# Transforming Student Evaluation with Adaptive Intelligence and Performance Analytics

Assistant Prof. Pushpalatha K S[1]
Dept. of ISE
Acharya Institute of Technology
Bengaluru (Affiliated to VTU, Belagavi),
Karnataka, India
pushpalatha23691@acharya.ac.in

Abhishek Mangalur[2]
Dept. of ISE
Acharya Institute of Technology
Bengaluru (Affiliated to VTU, Belagavi),
Karnataka, India
abhimangalur1@gmail.com

Ketan Hegde[3]
Dept. of ISE
Acharya Institute of Technology
Bengaluru (Affiliated to VTU, Belagavi),
Karnataka, India
ketanhegde2003@gmail.com

Chetan Badachi[4]
Dept. of ISE
Acharya Institute of Technology
Bengaluru (Affiliated to VTU, Belagavi),
Karnataka, India
badachichetan@gmail.com

Mohammad Aamir[5]
Dept. of ISE
Acharya Institute of Technology
Bengaluru (Affiliated to VTU, Belagavi),
Karnataka, India
mohdaamir4141@gmail.com

*Abstract* — **The development in Artificial Intelligence (AI) offers transformative potential for redefining student assessment methodologies. This paper aims to establish the idea of the advancement of Artificial Intelligence (AI) and its prospect in reshaping approaches to assessing students. It creates a system for the evaluation of students' performance using Artificial intelligence, and particularly the Gemini API for the generation of questions, grading and report on the student's performances. This is to facilitate easy use of the tools in creating, scheduling, and delivering assessments with minimal chances of cheating through options such as full screen and time limit. There are formats of questions in the system which comprises multiple choice, short answers and descriptive questions, developed by Gemini. The most conspicuous feature is the self-checking system whereby the user gets instant feedback for the correct score that each of the students would have scored instantly with explanations about wrong answers. Moreover, the platform has intelligent learning progressions where the user will be able to monitor his/her performances to be recommended a certain level of performance. It will allow students as well as educators to have real-time analytics and feedback on what they are good at and where they need to improve. Not only does it make the assessment easier, but it also improves the levels of accuracy in grading and effectively strengthens a data-based learning process for students.**

## I. INTRODUCTION

Traditionally, evaluation of students has been rooted in common practices that include preparation of questions, setting of examinations, grading as well as monitoring of student performance. However, they are labor intensive, time consuming and are vulnerable to development of human error [3]. For instance, there are common challenges as the consistency in grading, fairness in the assessment and the capacity to design an assessment that can meet the needs of every learner. In addition, the modern education is incredibly sophisticated to be evaluated by the existing methods in the present days.

Along the rapid advancement of Artificial Intelligence (AI) and machine learning technologies, there are learning opportunities that may reshape the educational processes. Informatics devices and programs that claim to do field work, provide individual recommendations or perspective from the vast big data. In other words, they establish this instrument as a useful solution to break the traditional student evaluation system constraints. Thanks to this, it is possible to create the environment of the adaptive, efficient and at the same time safe assessment of the results in which each participant would receive the benefit.

*A. Objective*

Develop an AI-powered student assessment system that would allow the implementation of the major assessment stages and would maintain student assessment integrity and information confidentiality at the highest level. The target areas are:

- Automated Question Generation
  Gemini generates many forms of questions, including multiple choice, short and significant questions, and descriptions among them. These questions therefore meet the curriculum objectives and the learning standards as per curriculum set by different institutions of learning.

- Exam Scheduling and Management
  It makes the process of managing the whole range of assessments easy and convenient for the educators, thus creating a smooth workflow [1].

- Automated Grading
  The system delivers consistent, accurate, and unbiased grading for different question formats. Students gets an instant and personalized feedback to support their learning process [4].

- Anti-Cheating Mechanisms
  To eliminate any cheating on the evaluation, the platform has enabled some measures such as full screen, browser lock, and time limit for the assessments.

Additionally, this research aims to integrate real-time performance analytics and adaptive learning pathways. The above facilitate the understanding of the progress, they also provide suitable suggestions as well facilitate lifelong learning for students. This ensures that educators can analyze the student performance and be able to make some useful changes to their strategies in a bid to enhance the teaching methodology.

*B. Significance*

This has a great prospect to compel educators to better focus on other higher-order tasks and responsibilities, in essence, curriculum enhancement, learners support, among other functions as some of the time-consuming routine tasks can be effectively addressed through such applications. The goal of this approach is to make the assessment fair and each student would be able to give their best performance to get the best result.

Additionally, the platform's integration of real-time performance analytics fosters a data-driven approach to education. Teachers can monitor the strengths, weaknesses and learning behavior of the students, which can be easily followed and differentiated between those fast learners and those that needs special focus. On the other side the platform also offers the accommodation in the context of students' continuous enhancement and students' self-motivated learning as well as the more creative and individualized educational approach for the students.

This work is an important step towards bridging the gap in modern learning ecosystem to enable proper utilization of resources for the evaluation techniques. The proposed platform would have benefits from AI, performance analytics, and that would make the process of assessing students more efficient and impactful, as well as equitable in the way the process of evaluating students is done.

II. BACKGROUND AND RELATED WORKS

*A. Development of a Web System with an Automated Question Generator Based on Large Language Models [1]*

This work proposes a web system for generating questions and answers based on LLM models made for several fields and where Serbian Wikipedia serves as the knowledge source. The idea of applying the LLM model in generating questions and answers makes it quite a unique way of creating quizzes. The incorporation of Wikipedia as the source of knowledge also enhances the quality of the generated questions. Not only can the web system generate questions for learners, but it also is capable of solving knowledge quizzes. A new set of questions for the quiz can be posted every one hour. Such functionality can help users to learn something new and enhance their skills in certain areas. Users can also keep track of their progress through response statistics. The web system can find its place in numerous platforms used for knowledge assessment. In general, this application brings together the LLM model, information from Wikipedia, and engaging quiz questions that enhances user learning.

*B. Using ChatGPT for Generating and Evaluating Online Tests [2]*

This paper aimed to assess the latest research on the quiz generation and its evaluation based on the ChatGPT-3 NLM model. From the literature research it was found out that the Quiz generation aspect is well discussed in literature since the launch of ChatGPT papers. It was established that the most significant issues were that although the model was developed and tested using a specific format of prompt, even minor changes to the format rendered the model wordy, unfocused and directionless about the task at hand. However, the results of the quiz were discussed in far fewer papers, and the primary focus was made on the summary production. The papers, which actually employed and experimented with NLPs for assessment, were few and highlighted the uncontrollability of the model because it changes the answer pattern even to similar questions, and the restrictive bias of the answers owing to the black box policy of the algorithm.

*C. A Comparative Analysis of Large Language Models with Retrieval-Augmented Generation-Based Question Answering System [3]*

This paper compares how efficiently Large Language Models (LLMs) are able to acquire and incorporate specific domain knowledge into their model. The Retrieval-Augmented Generation (RAG) system mitigates this by incorporating specific data, searching for relevant information from external knowledge resources, and improving the LLM's knowledge base for better responses. This research compared GPT-3.5 turbo, Gemini-Pro, and Llama-3 using the RAGAS evaluation framework in a RAG-based Question-Answering system. The evaluation focused on the performance comparison, where GPT-3.5-turbo stood out as the most effective model in using the retrieved data to enhance the answer quality, relevance, and correctness. Subsequent studies in this area might focus on improving the means by which information gets fed into LLMs with better filter and sort routines.

*D. Assessing the Use of OpenAI Chat-GPT in a University Department of Education [4]*

This paper elaborates on some issues related to the application of Chat GPT especially in a University Department of Education. A current investigation is underway in which the Department's students and its members will evaluate Chat GPT. Findings presented are based on a sample of an undergraduate course so they should be viewed as preliminary. It will therefore be useful to analyze tools like Chat-GPT in a bid to identify how they can be used bearing in mind the various negative outcomes they may entail. The dissemination of the results of the assessment will be useful to tool developers, education policymakers, teachers, students, and parents of minor students.

*E. How Teachers Can Use Large Language Models and Bloom's Taxonomy to Create Educational Quizzes [5]*

The purpose of this paper is to assert that it is possible for LLMs to generate various question types from a given context that would be just as effective as a written quiz that the teacher writes themselves. To achieve this objective, a series of quiz-writing experiments involving handwritten, simple, and controlled quizzes were undertaken. The controlled quizzes employed questions formulated to fit Bloom's taxonomy as a guide to categorization. Broadly and conclusively, the findings indicate that teachers exhibited a preference for writing quizzes with the use of controlled generations. They also mimic more of the controlled generations than the simple generations, meaning that these questions are of higher quality, or better for a teacher's interests. This supports our hypothesis that teachers benefit from automatically generated pedagogical questions for quiz development. Furthermore, the assessment of quiz quality revealed that the quizzes with controlled and simple generations are equally effective. Some are even indicative of their quality, especially when considered against handwritten quizzes. These findings will also guide the future of educational QG research in the direction that will benefit the targets and objectives of students and teachers.

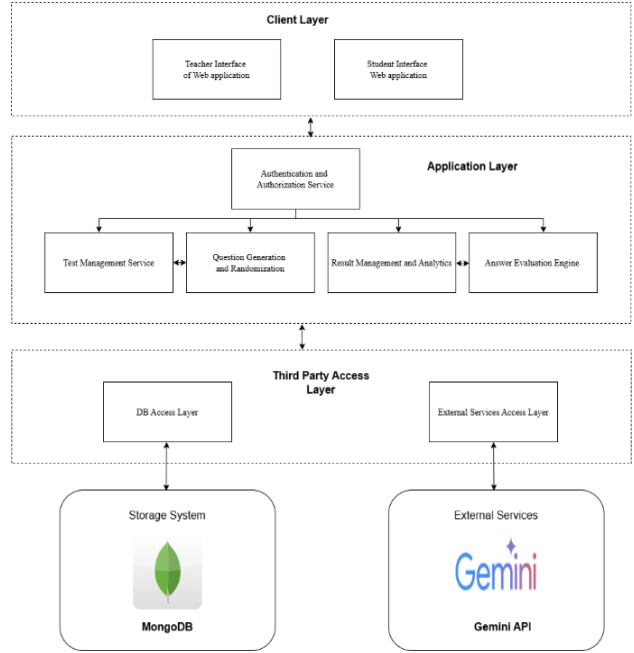

Fig. 1. System Architecture

## III. METHODOLOGY

The proposed system incorporates the Gemini API in enhancing the conventional student assessment system to be more effective, efficient, secure, and structured. The approach concerns itself with the major evaluation activities and structures that affect the performance of students whilst guaranteeing each learner an equitable process. The following components and methodologies are used in formulation and implementation of the system:

*A. AI-Driven Question Generation*

To meet curriculum standards and specific learning objectives, the platform leverages the Gemini API to automatically generate questions. These questions may consist of multiple-choice questions, short questions, and long questions that may be descriptive in nature which are commanded by the teachers. The topics aspect of the Gemini API guarantees that the questions used are diverse, relevant, and meet the assigned levels of difficulty to enable manufacturing effective assessment detail. The system also uses contextual analysis to formulate questions that bring critical analysis and other higher orders of questions such as problem solving. A dynamic aspect enables the variation of question level according to the identified ability level of students.

### B. Anti-Cheating Mechanisms

To ensure assessment integrity and fairness, multiple anti-cheating measures are implemented to ensure that none of the students engage in cheating. To enhance the security of student assessments full-screen mode, browser locking, and disabling keys and clicking are incorporated active features, and the use of time limits to avoid excessive student engagement with material is implemented as a passive feature. By implementing these features, it provides a trustworthy means to assess the outlined assessments.

### C. Performance Analytics

The platform augments the performance of the service through real-time performance tracking. With the use of predictive analysis, the system offers convenient outcome insights in the educational sector to the educators and students. It facilitates performance analysis that reveals students' strengths and all the aspects that require attention making it easy for the teachers to make the necessary decisions [2]. Students get such feedback along with the rate and pattern of performance analysis with directions for possible improvement for the specific student group. The analytics framework fosters continuous learning while also promoting transparency and accountability in the educational process.

### D. Automation and Efficiency

AI-driven automation facilitates the entire evaluation process, including question development, grading, and exam scheduling. The platform can help assist educators in the way the platform provides an algorithm that makes it easier for the teachers to create lessons, thus easing their burden. This puts the teachers in a better position to perform professional assignments such as improving students' motivation, and coming up with new appropriate strategies of assessing students with a view of improving performance Overall, the system allows for constant evaluations and causes no time lags as well as very few errors. Also, adaptive exam templates have been included to reflect on to the level of difficulty of the topic covered in the assessment as well as having the ability to be flexible in its assessment but at the same time being precise in its measures.

### E. Scalability

The platform is extensible with the ability to modify and expand it in accordance with the number of users and assessments. AI helps to optimally allocate resources thus; the system can effectively cope with several hundred of users. It is flexible which enables it to be applicable in all levels ranging from individual institutions to a country's different educational system. Furthermore, it has multilingual as well as regional adaptability that makes it useful in various learning environments. This makes the platform flexible in a way that it can accommodate other tests and introduce more domain subjects, to answer educational evolution.

### F. Security and Integrity

The assessments are structured to maintain standards of credibility and fairness for the tasks assigned to the techniques in the platform. Real-time monitoring, artificial intelligence in proctoring, along with behavioral analysis allow to identify and prevent cheating during exams. There are key security measures such as toggling the full-screen mode, the disabling of right-click, tab-closing and any other shortcut that may lead to test leakage in case someone tries to cheat. To ensure and uphold the integrity of the responses submitted, plagiarism check on the written answers is strengthened. These security measures enhance confidence and integrity in assessment to students as well as the educators who assess them.

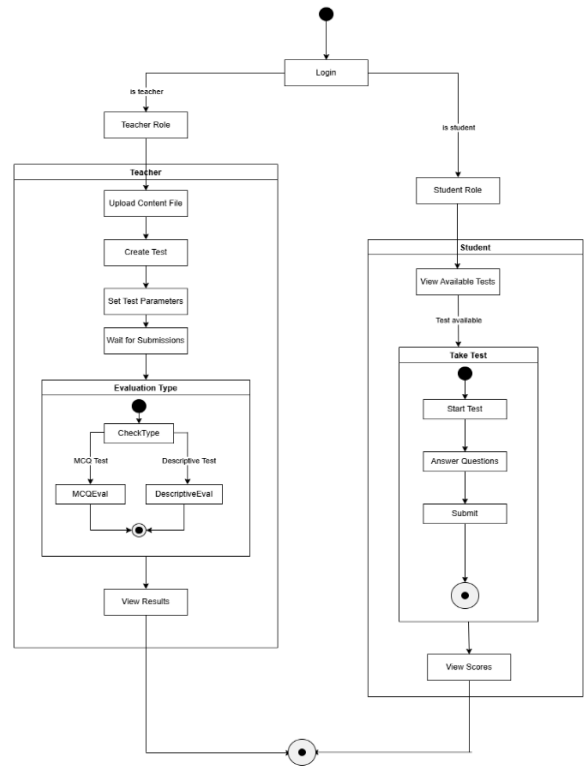

Fig. 2. Activity Diagram

## IV. IMPLEMENTATION

The implementation of the online assessment system involved integrating several modules to create a seamless and interactive platform for both teachers and students.

### A. Teacher Module

It allows teachers to contribute in the loading of content, creation of tests, setting of evaluation criteria and results. Some of the activities and features of the module are as follows:
- Login and Role Authentication: It can only be accessed by the Teachers through a secure login system to access administrative features.
- File upload: Teachers can upload educational content in various formats to support learning contents.
- Test creation: Test can be created as a type of dynamic form, one can select the type of the test (MCQ or descriptive), the number of questions and grading parameters.
- Setting Test Parameters: To maintain validity, the teachers can decide on the duration of an assessment, the time range within which the test must be completed.
- Evaluation Management: All the answers will be evaluated by the management suite of Gemini.

### B. Student module

This involves the student who can also be able to check available tests, take up tests, and check his/her scores. The process of the flowchart for the student module are as follows:
- Login and Role Authentication: Students log in securely to access the platform.
- View Available Tests: A list of tests for courses taken is provided to the students [1].
- Test Participation: Students can initiate a selected test within the stipulated time for the test.
- Result Viewing: It allows the students to see full analysis and outcome of the evaluation they have gone through.

### C. Backend Services

The backend of the system was implemented using Node.js with a MongoDB database for data storage.

### D. Evaluation and Security Features

The four levels of security features, which allow the achievement of more reliable results of examination, are Full-Screen mode, Tab-Switch detector, Test Submission automation and data protection [2].

## V. EVALUATION PROCESS

The evaluation of the online assessment system involves the use of the Gemini API to facilitate the correct assessment of the MCQs as well as the scoring of the descriptive questions. It is necessary to note that this method of evaluation is fully automated and can be deployed in the real-time mode, while utilizing AI technologies.

### A. Automated MCQ Evaluation

The evaluation of MCQ questions is conducted entirely by the Gemini API. The process follows these steps:
- Answer Matching: The Gemini matches student answers against the correct options.
- Automated Grading: All the respective questions carry full marks if a student gets the correct answer. In cases of an incorrect or unanswered question, no marks are awarded (or negative marking, if configured).
- Instant Results: The Gemini results instantly and returns the total scores of the students, rank and time taken among others.
- Performance insights: The teachers as well as students both receive detailed feedback, including response accuracy and performance pattern.

### B. Automated Descriptive Answer Evaluation

Descriptive answers, which traditionally require human evaluation, are assessed using the Gemini API's natural language processing (NLP) capabilities. The evaluation involves:
- Semantic Analysis: The Gemini API provides the analysis of the student responses and the comparison to the pre-defined rubrics with parameters: relevance, Grammar, Coherence, Completeness, and depth [5].
- Scoring Logic: The Gemini evaluates responses simply by matching the responses given with one word or a set of words given to the ideal responses.
- AI-Powered Feedback: The Gemini generated automated feedback to help students understands areas for improvement.

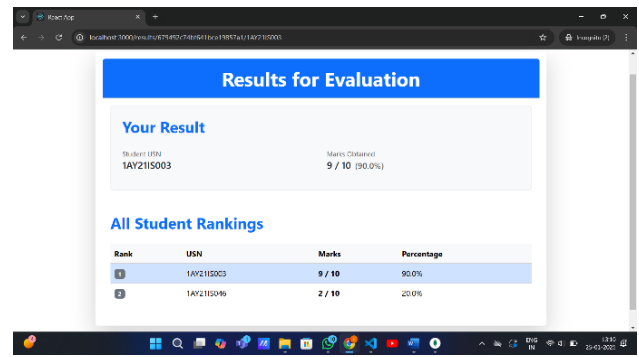

Fig. 3. Detail Report

The Fig. 3. displays above show the results of an online evaluation. It shows the user's individual score, 9/10 or 90%, along with their rank among other students. The table below lists the rankings of all participants, including their USN (University Seat Number), marks, and percentage.

## VI. RESULTS AND DISCUSSION

TABLE 1. Performance Metrics

| Metric | Value |
|---|---|
| Question Accuracy | 98% |
| Grading Consistency | 97% |
| Feedback Timeliness | Instant |

### A. Evaluation Efficiency

The platform's use of the Gemini API for automated evaluation significantly improved the assessment process. Multiple choice questions, short answer and descriptive question assessments were done with high reliability and comparability. This led to the minimizing of the waiting time that students used to spend waiting for results from the instructors as the system offered instant results to the students.

### B. User Feedback

Students elaborated on self-test data that stated more instant data response and made the participants know their mistake and area of deficiency after submitting their tests. According to experiences of the teachers some of the features like auto scoring and fewer concerns of grading on paper was seen as good.

### C. Adaptability Success

The platform has been designed in such a way that it fit perfectly with the existing educational processes and Learning Management Systems (LMS). Thus, its versatility in covering various subject fields and kinds of tests proved useful for various educational establishments, including small schools and networks.

### D. Security Analysis

Strategies such as behavioral monitoring in real time as well as automated test monitoring worked to counter cheating. The secure and transparent evaluation system-built trust among both students and educators [4].

### E. Scalability Results

The system's architecture demonstrated robust scalability during stress tests, efficiently handling up to thousands of concurrent users with stable performance. It also included support for multilingual content and broadened its use in the educational institutions.

## VII. CONCLUSION

This research adopts the Gemini API and developing an automated assessment system for multiple choice and descriptive kind of questions. It also simplifies the assessment process by eliminating the possibility of following the wrong answer paths as well as provides a means of automatically generating the results in every assessment occasion.

The conclusion of the proposed solution captures the major problems facing modern educational assessment and they include a time-consuming process of manual assessment and subjective grading. The flexibility of the system can easily fit into current education setups and accommodate multiple curricula as well as various and methods of evaluating students.

The results shows that evaluation efficiency has improved along with the reliability of the assessment with benefits that make the platform more applicable to large-scale educational application. Further development of the program will involve the improvement of the evaluation capacity, multi-lingual features, and, consequently, the presence of diversified subject areas required by learners.